\documentclass[aps,pre,reprint,showpacs,superscriptaddress]{revtex4-1}
\usepackage{graphicx}
\usepackage{amsmath}
\usepackage{epstopdf}

\usepackage{color}

\begin{document}
\title{Nonlinear Time-Reversal of Classical Waves: Experiment and Model }
\author{Matthew Frazier}
\affiliation{Department of Physics, University of Maryland, College Park, MD  20742-4111}
\author{Biniyam Taddese}
\author{Bo Xiao}
\author{Thomas Antonsen}
\author{Edward Ott}
\author{Steven M. Anlage}
\affiliation{Department of Physics, University of Maryland, College Park, MD  20742-4111}
\affiliation{Department of Electrical and Computer Engineering, University of Maryland, College Park, MD  20742-3285}
\date{\today}

\begin{abstract}
We consider time-reversal of electromagnetic waves in a closed, wave-chaotic system containing a discrete, passive, harmonic-generating nonlinearity. An experimental system is constructed as a time-reversal mirror, in which excitations generated by the nonlinearity are gathered, time-reversed, transmitted, and directed exclusively to the location of the nonlinearity. Here we show that such nonlinear objects can be purely passive (as opposed to the active nonlinearities used in previous work), and develop a higher data rate exclusive communication system based on nonlinear time-reversal. A model of the experimental system is developed, using a star-graph network of transmission lines, with one of the lines terminated by a model diode. The model simulates time-reversal of linear and nonlinear signals, demonstrates features seen in the experimental system,  and supports our interpretation of the experimental results.
\end{abstract}


\maketitle

\section{Introduction} 

The time-reversal and reciprocal properties of the lossless linear wave equation can be utilized to achieve useful effects even in wave-chaotic systems\cite{a1} typically endowed with complex boundaries and inhomogeneities.\cite{a2,a3,a4,a5,a6,a7,a8,a9} Wave equations without dissipation are invariant under time-reversal; given any time-forward solution considered as a superposition of travelling waves, there exists a corresponding time-reversed solution in which the individual superposed travelling waves propagate backwards retracing the trajectories of the time forward solution. In principle, this allows the construction of a time-reversal mirror.  First, imagine an ideal situation in which one transmits a waveform of finite duration from a localized source in the presence of perfectly reflecting objects and then receives the resulting reverberating waveforms (referred to as the {\it sona}) on an array of ideal receivers completely enclosing the region where the source and reflecting objects are located. After the reverberations die out, one then transmits (in the opposite direction) the time-reversed sona signals from the array of receivers.  This newly transmitted set of signals essentially undoes the time-forward propagation, producing waves which converge on the original localized source, reconstructing a time-reversed version of the original signal at the localized source. Although real situations deviate from the above described ideal, time-reversal in this manner has been effectively realized in acoustic \cite{a2,a3,a4,a5,a6,a7,a8,a9, a11, a12} and electromagnetic waves \cite{a6, a8, a13, b13}, and applications such as lithotripsy \cite{a2, a4}, underwater communication \cite{a2, a14, a15}, sensing small perturbations \cite{a11, a12}, and achieving sub-wavelength imaging \cite{a6,a7,a8, a16} have been developed.

Ideally, for a perfect time-reversal mirror, a large number of receivers are required to collect the sona signals, and the receivers need to cover a surface completely surrounding the source and any reflecting objects (which reflect without loss). A significant simplification is to enclose the system in a closed, ray-chaotic enviroment with highly reflecting boundaries. For wavelengths smaller than the enclosure size, propagating waves will (over a sufficiently long duration) reach every point in the enviroment, allowing a single wave-absorbing receiver to record a single time-reversible sona signal over a long duration.\cite{a9,a11}  Somewhat surprisingly, it has been found that, in the presence of boundary reflection loss, a high-quality version of the basic time-reversal reconstruction still occurs at the source, and reception of only a small fraction of the transmitted energy is sufficent for reconstruction of the initial waveform at the source. Nevertheless, such a time reversal mirror still requires an active source to generate the sona signal. In some cases, it would be better if this step could be eliminated, further simplifying the time reversal mirror.

Recent studies have investigated the addition of discrete elements with complex nonlinear dynamics to otherwise linear wave-chaotic systems. \cite{a10, arxiv1} When a discrete nonlinear element is added to the system, excitations at new distinct frequencies are generated from the interaction of the initial waveform with the element. This appears as a radiated signal originating at the location of the nonlinear element (which in principle may be unknown). The new waveform propagates through a linear medium, and when time-reversed and retransmitted, will reconstruct the excitations generated at the nonlinear element. (This form of nonlinear time-reversal differs from wave propagation through a distributed nonlinear medium, in which the time-reversal invariance breaks when shock waves form.\cite{a19}) Time reversal in systems with localized nonlinearities has been demonstrated in several systems: acoustic waves through materials with defects \cite{a17}, as a means of non-destructive evaluation\cite{a18}, phase conjugation of light harmonically generated from a nanoparticle \cite{a21}, and phase conjugation of acoustically modulated light using a focused utrasonic signal as a 'guide star' for the time-reversed focusing\cite{a20}.

In our previous work, nonlinear time-reversal was performed using microwaves incident upon a harmonically-driven diode, generating intermodulation products. \cite{arxiv1}. A drawback of the technique is the need to use an active nonlinearity (a driven diode to create intermodulation products) instead of a passive element. Also missing is a quantitative model to describe and understand the nonlinear time-reversal physics. Furthermore, the technique was applied to develop a method of exclusive communication, which appears to be rate-limited by the length of the sonas necessary to transmit information.  Here we examine a wave chaotic system with a discrete, passive nonlinear element as a nonlinear time-reversal mirror, and construct a model system using a star graph network of transmission lines to simulate propagation through this nonlinear wave-chaotic system. We also demonstrate a method of transmitting information encoded in the nonlinear sona, explore the extent to which the recorded length of the sona limits the rate of information transfer, and demonstrate the ability to reconstruct overlapped sonas into distinct pulses, allowing for compressing many bits of information into a given sona length.


\section{Experiment}
\begin{figure*} 
\includegraphics[width=0.8\textwidth]{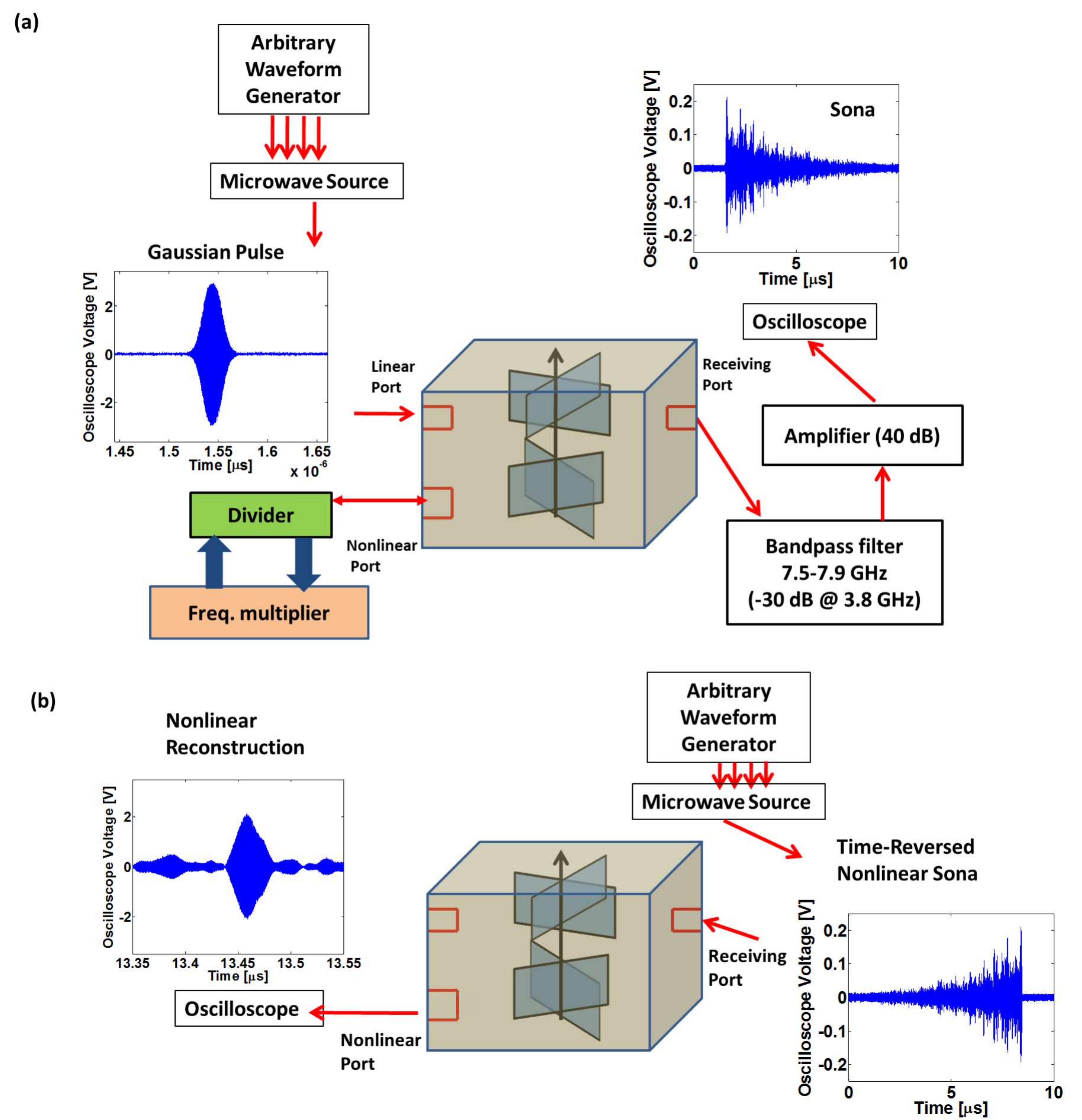}
\caption{(Color online) (a) The experimental setup in the time-forward step, containing a passive nonlinear circuit, showing the Gaussian pulse injected into the linear port and the resulting sona measured at the receiving port. (b) The experimental setup in the time-reversed step, showing the time-reversed nonlinear sona injected in to the receiving port and reconstructing as a time-reversed excitation at the nonlinear port.  } \label{fig:NewExpt}
\end{figure*}

\begin{figure*} 
\includegraphics[width=0.8\textwidth]{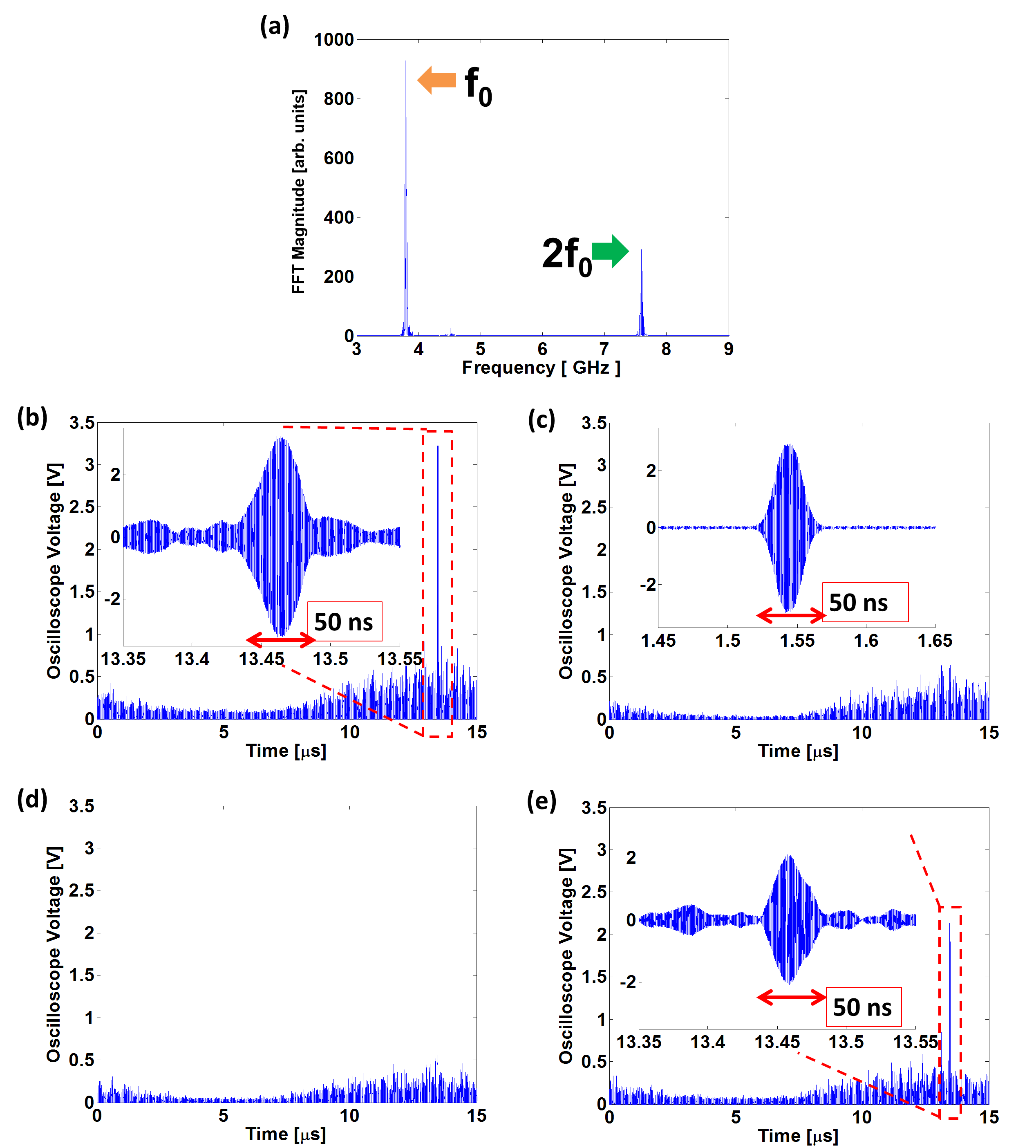}
\caption{(Color online) (a) Fourier transform of the sona measured at the receiving port, indicating the linear sona at $f_{0}$ and nonlinear sona at $2f_{0}$. (b) Reconstruction of the linear sona at the linear port. (c) There is no appreciable reconstruction of the nonlinear sona at the linear port (inset shows the original incident pulse, for comparison). (d) There is no appreciable reconstruction of the linear sona at the nonlinear port. (e) Reconstruction of the nonlinear sona at the nonlinear port.  } \label{fig:a5}
\end{figure*}

We have realized a time-reversal mirror using electromagnetic waves in a closed complex (ray-chaotic) scattering environment.    Figure \ref{fig:NewExpt}(a) illustrates the set up showing a Gaussian pulse injected at the linear port and the sona measured at the receiving port.  The enclosure is a 1.06 $m^3$ aluminum box ($ 1.26 m \times 1.26 m \times 0.67 m$) with irregular surfaces and a conducting scattering paddle, and has three ports, consisting of loop antennas, for the introduction and extraction of microwave signals.  A passive nonlinear element was introduced by connecting one port to a frequency multiplier circuit.  The circuit consists of a Wilkinson divider (HP model 87304C), with both outputs connected to the ports of a passive diode-based x2 frequency multiplier (Mini-Circuits model ZX90-2-50-S+) to form a closed circulating nonlinear circuit.  The circuit is configured to mimic the response of a memoryless, passive, nonlinear element.  In the time-forward portion of the experiment, an initial driving signal, consisting of a Gaussian-shaped pulse at a carrier frequency $f_{0} =$ 3.8 GHz is transmitted into the system from the linear port (Fig. 1(a) inset). The excitation propagates throughout the system, including to the nonlinear element (60 cm from the linear port), where the portion of the Gaussian pulses incident upon this circuit is partially upconverted to twice the pulse frequency ($2 \times f_{0} =$ 7.6 GHz) before re-entering the enclosure with a time delay of less than 1 ns. The combined sona  signal [time-domain shown in Fig. \ref{fig:NewExpt}(a), frequency domain in Fig \ref{fig:a5}(a)] is received at the receiving port (125 cm from the linear port, 150 cm from the nonlinear element) and filtered in software into a ``linear sona'' at the pulse carrier frequency, and a ``nonlinear sona'' at the second harmonic frequency. (When the frequency multiplier circuit is removed from the system, the second harmonic disappears from the received sona signal, as expected, leaving only the ``linear sona''.)  

In the time-reversed portion of the experiment (Fig. \ref{fig:NewExpt}(b)), each sona is time-reversed and retransmitted at the power level of the original pulse from the receiving port.  The resulting signals are then measured at the linear port and at the nonlinear port (the nonlinear circuit is removed in this step). Figure \ref{fig:a5} (b-e) shows example reconstructions measured using this setup. For the linear sona, a reconstructed Gaussian pulse appears only at the linear port (Fig. \ref{fig:a5} (b)); similarly, for the nonlinear sona, a reconstructed pulse appears only at the nonlinear port (Fig. \ref{fig:a5} (e)). This demonstrates the exclusive nature of information transfer between the receiving port and the linear / nonlinear port.\cite{arxiv1}  As explained previously\cite{arxiv1}, and examined further below, we interperet the reconstructed waveform in Fig. \ref{fig:a5}(e) as that arising from the first large-amplitude pulse that arrived at the diode during the time forward phase of the experiment (Fig. \ref{fig:NewExpt}(a)). 

\begin{figure*} 
\includegraphics[width=0.65\textwidth]{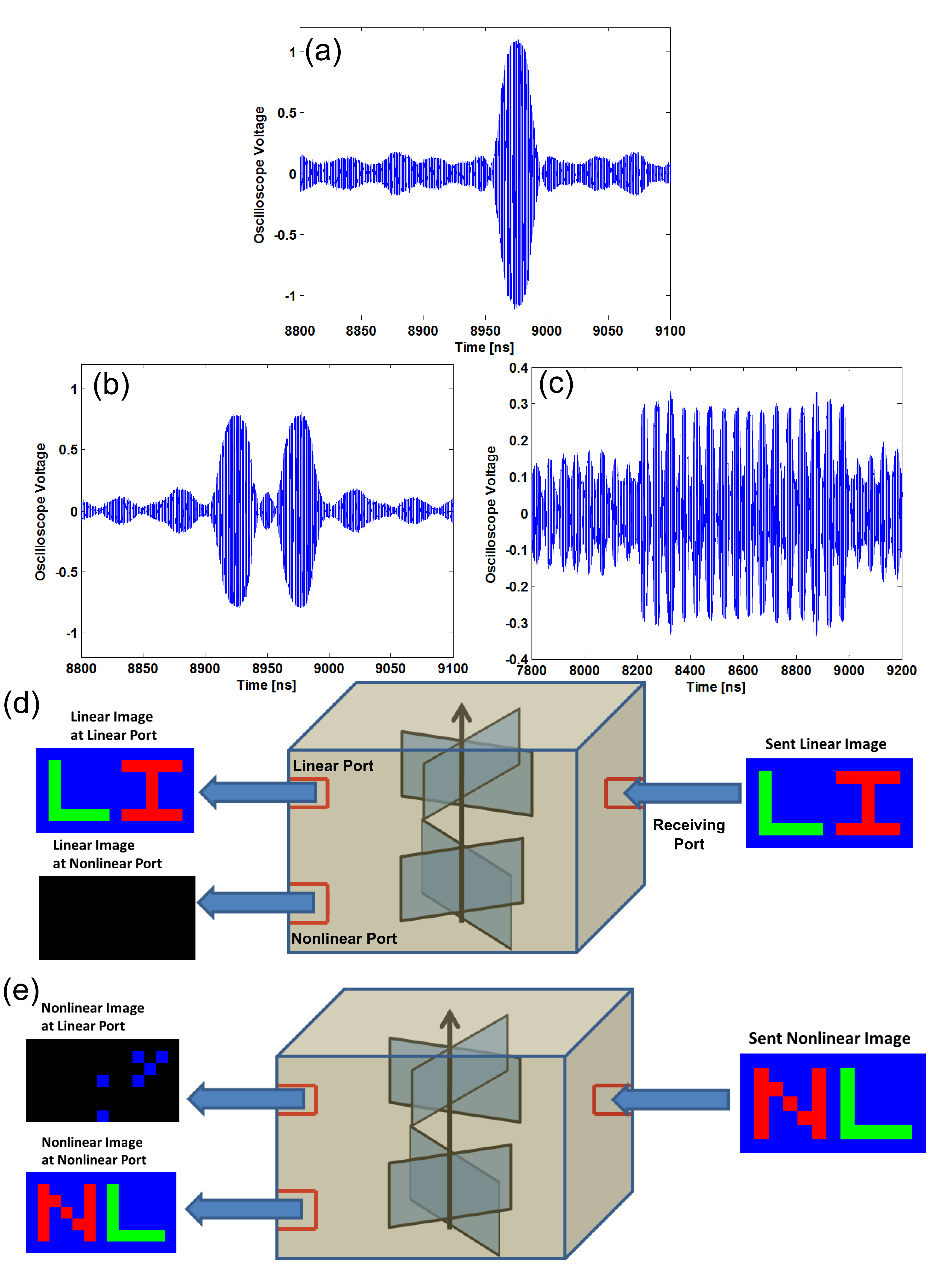}
\caption{(Color online) (a) Reconstruction of a single linear time-reversed sona at the linear port generated using a 50ns Gaussian pulse. (b) Reconstruction of a sona at the linear port constructed by overlapping two copies of the sona with a 50 ns time delay (or 99.5\% overlap of sonas), showing two distinct reconstructed pulses of lower amplitude. (c) Reconstruction of a sona at the linear port constructed using sixteen overlapped sonas, with 50 ns time delays between copies. Sixteen reconstructed pulses are distinguishable above the noise / sidelobes. (d-e) Images of ``LI'' and ``NL'' encoded using the linear sona (d) and nonlinear sona (e), transmitted exclusively to the linear and nonlinear ports in the experimental setup, respectively.}
\label{fig:images}
\end{figure*} 

The exclusive nature of the reconstruction of the nonlinear sona allows the construction of a communication channel to the nonlinear port.\cite{arxiv1}  Figure \ref{fig:images} demonstrates such a communication channel in the experimental system using the passive harmonic-generating nonlinearity. The communication channel works by using a signal formed by shifting and superposing a series of one-input-pulse sonas.  Figures \ref{fig:images}(a-c) demonstrate reconstructions of the overlapped sonas. When this superposition of sonas is time reversed and reinjected, it results in reconstructing a sequence of pulses at the intended receiver.  To maximize the data capacity it is desired to employ shifts that are relatively short, yet long enough that the reconstruction results in distinguishable pulses.  In Fig. \ref{fig:images}(a), a reconstruction from a single linear sona is shown for illustration. In Fig. \ref{fig:images}(b), two sonas are overlapped, with a time delay of 50 ns (shifted one pulse duration \cite{pulseoverlap}). The two pulses are clearly resolved in the reconstruction, though they are of lower amplitude than in the single-pulse reconstruction in Fig. \ref{fig:images}(a); the combined sonas are transmitted at the same power as the single sona, which results in the energy division between the reconstructions. In Fig. \ref{fig:images}(c), sixteen sonas are combined, each new sona shifted 50 ns from the previous one. The reconstructions are still distinguishable above the sidelobes, which are somewhat enhanced.

To send binary coded information as a string of 1's and 0's, we use a transmitted time-reversed one-pulse sona to represent a 1 and a phase-scrambled version of the one-pulse sona to represent a 0 (Ref. \cite{arxiv1}).  The use of a phase-scrambled sona for a 0 (rather than no transmission at all for a 0) is advantageous in that an eavesdropper, at another location, would not be able to distinguish 0's from 1's.  Figures \ref{fig:images} (d) and (e) show images transmitted using reconstructions of concatenated sonas; each pixel color is encoded as a two-bit word ( black = '00', blue = '01', red = '10', green = '11') in a sona overlapped by 50\% (a time delay of 5 $\mu$s). The ``LI'' image in (d) was encoded using the filtered \textit{linear} sona and transmitted from the receiving port. The reconstructions at the linear port were decoded into pixels of the appropriate color, generating a facsimilie of the original image. At the nonlinear port, no reconstruction arrived, and the decoded waveform (a series of '00' words) decoded as a black image. The ``NL'' image shown in (e) was encoded using the filtered \textit{nonlinear} sona and transmitted in the same manner from the receiving port; here, the reconstructions arrive at the nonlinear port, and decode as the facsimilie image. At the linear port, the lack of reconstructions result in a mostly black image.
\section{A model for nonlinear time-reversal}

\subsection{ Linear Model }

\begin{figure*}
\includegraphics[width=\textwidth]{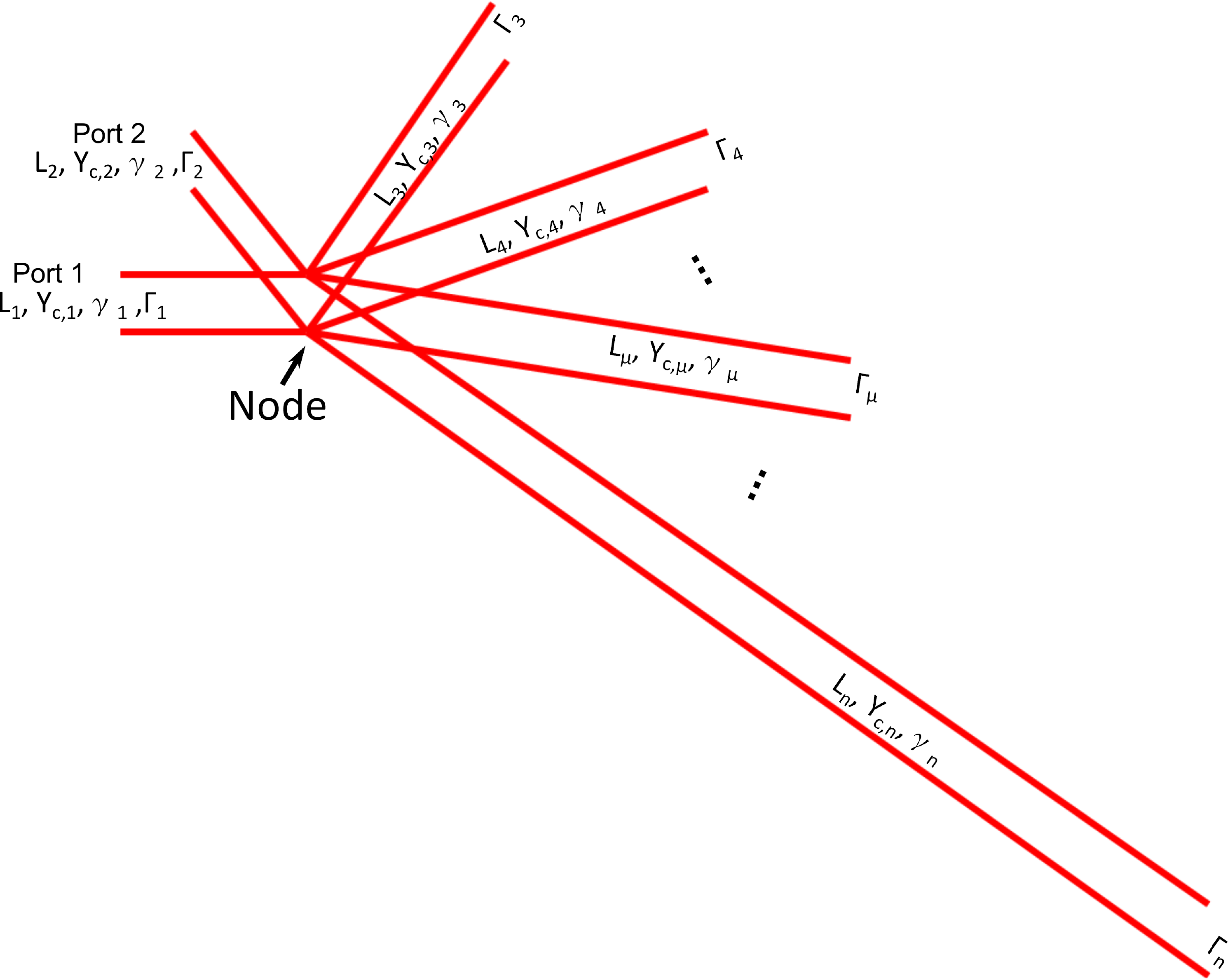} 
\caption{(Color online) The model network of transmission lines connected in a star graph topology. For each line, a length ($L_\mu$), characteristic admittance ($Y_{c,\mu}$), and reflectivity ($\Gamma_\mu$) of the unconnected end is defined. Ports 1 and 2 act as the points where energy is injected into, or extraced from, the graph.}\label{fig:1a}
\end{figure*}

Our model of a cavity with two ports is constructed as a network of transmission lines of varying lengths, connected in a star graph topology,\cite{b13,c13,KSG} as shown in Fig. \ref{fig:1a}. The use of multiply-connected transmission lines to simulate complex wave-chaotic scattering systems is well established \cite{KSG}. Wave propagation on a network of inter-connected transmission lines becomes quite complex, even for a simple graph. The computational simplicity of these quasi-one-dimensional models adds to their appeal. Each pair of lines in Fig.\ref{fig:1a} depicts a linear homogeneous transmission line, representing a portion of a channel of wave propagation in the cavity.  Each line $\mu$ is assigned a length $L_\mu$ (and a corresponding time for a wave packet to transit $\Delta t_\mu$), a characteristic admittance $Y_{c,\mu}$, attenuation ($\alpha_\mu$) and phase evolution ($\beta_\mu \equiv \omega / v_\mu$) per unit length (with combined propagation constant $\gamma_\mu(\omega) = \alpha_\mu + i \beta_\mu$), and a complex reflectivity $\Gamma_\mu$ for the termination of the unconnected end. Lines 1 and 2 are designated to represent the two ports into the cavity by selecting $\Gamma_1 = \Gamma_2 = 0$; waves exiting the graph through the two ports are not reflected.  The other lines $\mu > 2$, through the multiple reflections and scattering at the common node, simulate the presence of multiple ray paths connecting ports 1 and 2.  In what follows, we shall henceforth consider the case $\Gamma_{\mu} = 1$ (open circuit) for $\mu > 2$. (In previous models, demonstrated in \cite{b13,c13}, only a single port is used to transmit and receive waveforms.) A waveform is ``injected" into the system at one port, and the resulting waveform is ``received" and recorded at the other port.

The linear transmission line networks are analytically solvable in both the frequency domain and in the time domain. In the frequency domain, a sona is generated by applying the scattering matrix to an initial Gaussian pulse. For a two-port star network of transmission lines, the scattering parameters can be expressed in terms of the input admittance ($Y(\omega)$) of the (non-port) lines and the characteristic impedances of the ports ($Z_{c,\mu} \equiv 1 / Y_{c,\mu}$) by 

\begin{subequations}\label{eq:Smatrix}
  \begin{align}
Y(\omega) &= \sum_{\mu=3}^{n} Y_{c,\mu} \left ( \frac{ 1 - \Gamma_\mu e^{-2\gamma_\mu L_\mu}}{1 + \Gamma_\mu e^{-2\gamma_\mu L_\mu}} \right ) \\
S_{11}(\omega) &= \frac{Z_{c,2} - Z_{c,1} - Y(\omega) Z_{c,1} Z_{c,2}}{{Z_{c,1} + Z_{c,2} + Y(\omega)Z_{c,1} Z_{c,2}} } e^{-2\gamma_1 L_1}\\
S_{12}(\omega) &= \frac{2 \sqrt{Z_{c,1} Z_{c,2}}}{{Z_{c,1} + Z_{c,2} + Y(\omega)Z_{c,1} Z_{c,2}}}  e^{-\gamma_1 L_1 - \gamma_2 L_2}\\
S_{21}(\omega) &= S_{12}(\omega)\\
S_{22}(\omega) &= \frac{Z_{c,1} - Z_{c,2} - Y(\omega) Z_{c,1} Z_{c,2}}{{Z_{c,1} + Z_{c,2} + Y(\omega)Z_{c,1} Z_{c,2}}}e^{-2\gamma_2 L_2}
\end{align}
\end{subequations}
where $n$ is the total number of lines. We choose to make port 1 perfectly matched (i.e., no prompt reflection for signals injected into this port) by requiring
  \begin{align}\label{eq:matching}
Y_{c,1} &=  \sum_{\mu=2}^{n} Y_{c,\mu} = Y_{c,2} + \sum_{\mu=3}^{n} Y_{c,\mu}
\end{align}
From Eq. (\ref{eq:matching}) it is not possible to simultaneously match both port 1 and port 2.\cite{sameer}  In what follows, we choose $Y_{c,2} = Y_{c,1}/2$, for which there is a prompt reflection coefficient of $1/3$ at port 2.

In the time-domain model, a waveform is 'injected' from one port into the network. At a particular time $t$, the voltages in the network are described by the following system of equations. The voltage $V_{N}(t)$ at the node connecting the transmission lines in parallel is given by 
\begin{equation}\label{eq:a}
  V_{N}(t) = (V_{\mu, +} + V_{\mu,-}), 
\end{equation}
for all $\mu$, where $V_{\mu,+}$ represents the voltage of the incoming wave along line $\mu$ at the node, and $V_{\mu,-}$ the voltage of the outgoing wave. The current entering the node contributed by line $\mu$ is 
\begin{equation}\label{eq:b}
  I_\mu(t) = Y_{c,\mu} \left [ V_{\mu,+} - V_{\mu,-} \right ]
\end{equation}
Summing over all n lines in the network,
\begin{equation}\label{eq:c}
  \sum_{\mu=1}^{n} I_\mu(t) = 0
\end{equation}
as required by Kirchoff's current law. Finally, the voltage for a wave reflected from a line end and incoming to the node is given by
\begin{equation}\label{eq:d}
  V_{\mu,+}(t) = e^{-2\alpha_\mu L_\mu} V_{\mu,-} \left (t - 2\Delta t_\mu \right )
\end{equation}
where $e^{-2\alpha_\mu L_\mu}$ accounts for the attenuation, and the reflection coefficient at the end of the line is taken to be $\Gamma_{\mu} = 1$ ($\mu > 2$), and the attenuation $\alpha_{\mu}$ has been approximated as being constant over the bandwidth of the signal. 

Using Eq. (\ref{eq:a}) to express $V_{\mu,-}$ in Eq. (\ref{eq:b}) and substituting in Eq. (\ref{eq:c}) determines the node voltage $V_{N}$ in terms of the incoming voltages $V_{\mu,+}$, 
\begin{equation}\label{eq:e}
  V_{N}(t) = 2 \frac{\sum_{\mu=1}^{n} Y_{c,\mu}  V_{\mu, +} }{\sum_{\mu=1}^{n} Y_{c,\mu}}.
\end{equation}

A sona is calculated from the initial Gaussian pulse, propagated from the broadcasting port (port 1) as $V_{1, +}(t)$, by discretizing time, expressing $V_{\mu, -}=V_{N}-V_{\mu, +}$, and using Eq. (\ref{eq:d}) to update $V_{\mu, +}$.  Once $V_{N}(t)$ is known, the waveform arriving at the receiving port can be recorded. In the time-reversed direction, the transmitting and receiving ports are swapped, and time-reversed sona is propagated from the new transmitting port, creating a reconstruction at the new receiving port. 

\begin{figure*}     
\includegraphics[width=\textwidth]{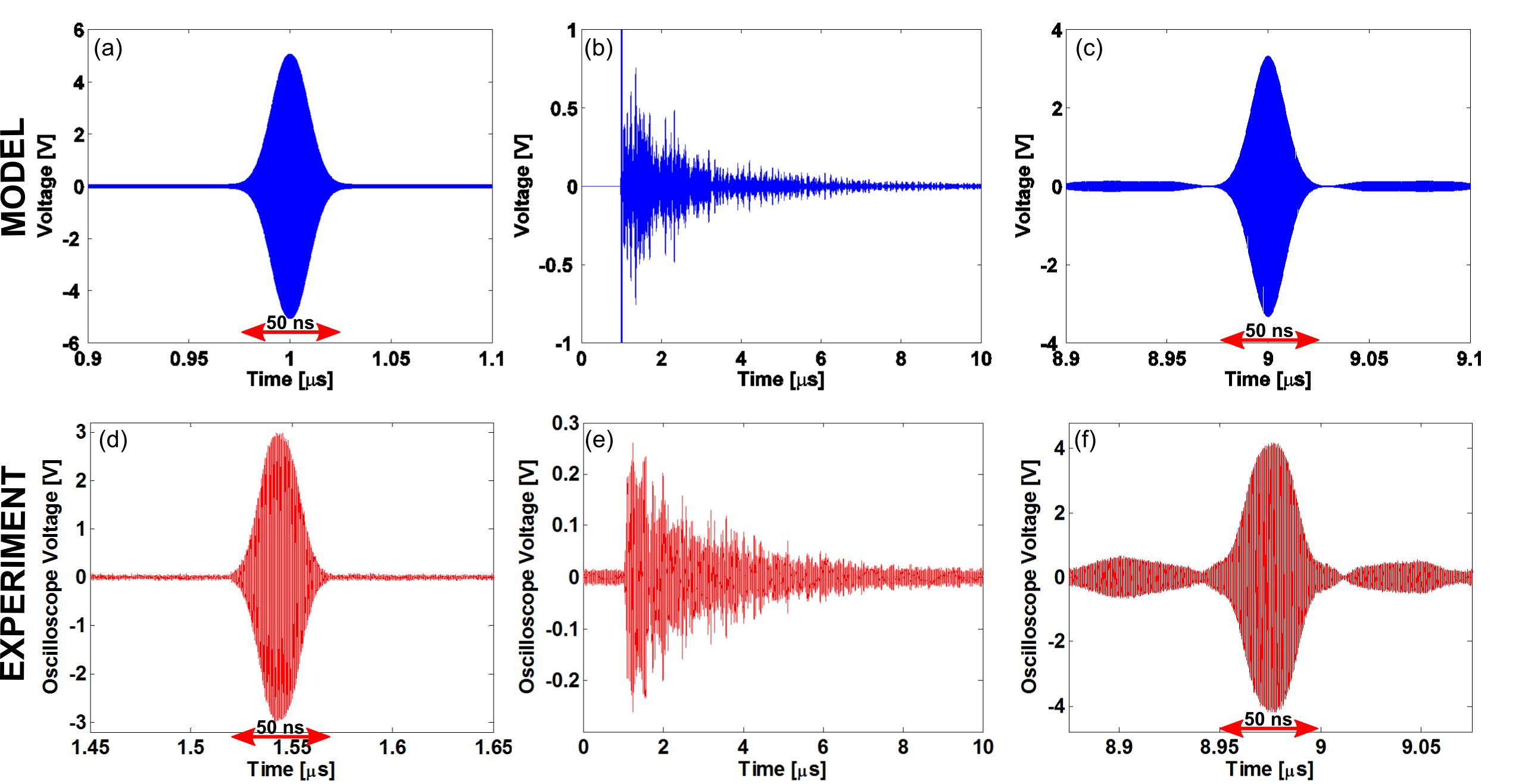}
\caption{(Color online) (a-c) Time domain signals in the linear 2-port star graph with n=10 lines. (a) Initial Gaussian pulse (50ns, 3.8 GHz carrier) applied to port 1. (b) Resultant sona collected at port 2. (c) Reconstruction at port 1 from linear time-reversal of sona in (b) generated in the model. (d-f) Experimental results obtained in the linear resonant cavity. (d) Initial Gaussian pulse (50 ns, 3.8 GHz carrier) applied to the transmitting port. (e) Resultant sona collected at the receiving port. (f) Reconstruction of the time-reversed sona in (e) recorded at the original transmitting port.} \label{fig:1bc}
\end{figure*}

Figure \ref{fig:1bc} shows the sonas and reconstructions generated by the described model (a-c) and the linear time-reversal experiment (d-f), for comparison. The sona in Fig. \ref{fig:1bc}(b) was generated by propagating the Gaussian pulse with carrier frequency $f_{0} =$ 3.8 GHz shown in (a) through a transmission line network from port 1 to port 2. The network was constructed from ten line segments of  lengths ($1, 1, 4, 6, 10, 16, 25, 40, 60, 100 $)m \cite{stargraph}, with no loss ($\alpha_\mu = 0$), and open circuited ends ($\Gamma_\mu = 1$) for the non-port lines. The characteristic admittances are $Y_{c,\mu} = 0.02 $ S for lines 3-10, $Y_{c,2} = 0.16$ S  and $Y_{c,1} = 0.32$ S (see Eq. (\ref{eq:matching})). Note that the decay of the sona signal in Fig. \ref{fig:1bc}(b) is due to power leakage out through ports 1 and 2. The reconstruction at port 1 shown in Fig. \ref{fig:1bc}(c) was generated by propagating the time-reversal of the sona into the network from port 2.

In the experiment, the Gaussian pulse at carrier frequency $f_{0} =$ 3.8 GHz shown in (d) was transmitted into the three dimensional linear cavity at the transmitting port, generating the sona (e) recorded at the receiving port. The time-reversed sona was retransmitted into the cavity at the receiving port, reconstructing the waveform shown in (f) at the transmitting port. The model closely approximates the real time-reversal mirror, as shown by the sidelobes ($t \lesssim 8.975 \mu$ s, $t \gtrsim 9.025 \mu$ s) in the reconstruction in (c), which are similar to those seen in (f). The model sidelobes result because perfect time reversal is not achieved because the time-forward reflected signal propagating backward in line 1 is not recorded, reversed and reinjected.  Also, a portion of the sona signal is lost due to the finite recording time in the time-forward step, further degrading the reconstruction.  In addition to these mechanisms, the sidelobes in the experiment arise from attenuation of the signals as they propagate. 

\subsection{ Diode Model }
The nonlinear element placed in the experimental cavity is modeled as a diode terminating one transmission line, with incoming waveforms reflecting off of the termination. The diode introduces harmonics to the reflected waveform, and does not model any particular diode in detail. The model also lacks a time-delay or history-dependent mechanism; there is no ``memory'' of previous states, as in more sophisticated diode models.\cite{ren1}  In terms of the incoming and outgoing voltages, the voltage across and current through the diode is given by 
\begin{align}\label{eq:da}
  V_d &= V_+ + V_-\\\label{eq:db}
  I_d &= Y_c [ V_+ - V_- ]
\end{align}
where $Y_c$ is the characteristic admittance of the transmission line connected to the diode. 
The current through the diode is also expressed in terms of the diode voltage by the ideal diode equation:
\begin{equation}\label{eq:dc}
  I_d = I_s (\exp{\frac{V_d}{V_T} -1}),
\end{equation}
where $I_s$ is the saturation current of the diode, and $V_T = kT / e$ the thermal voltage. Solving this system of equations for $V_-$ in terms of $V_+$ ( and defining $f \equiv \frac{Is}{Y_cV_T}$) gives

\begin{equation}\label{eq:dsol}
  V_- = V_+ + \frac{I_s}{Y_c} - V_T \ W( f e^{2\frac{V_+}{V_T} + f} ) 
\end{equation} 
where $W(z)$ is the Lambert W-function, defined as the inverse of $z = We^W$.
\begin{figure*}     
\includegraphics[width=\textwidth]{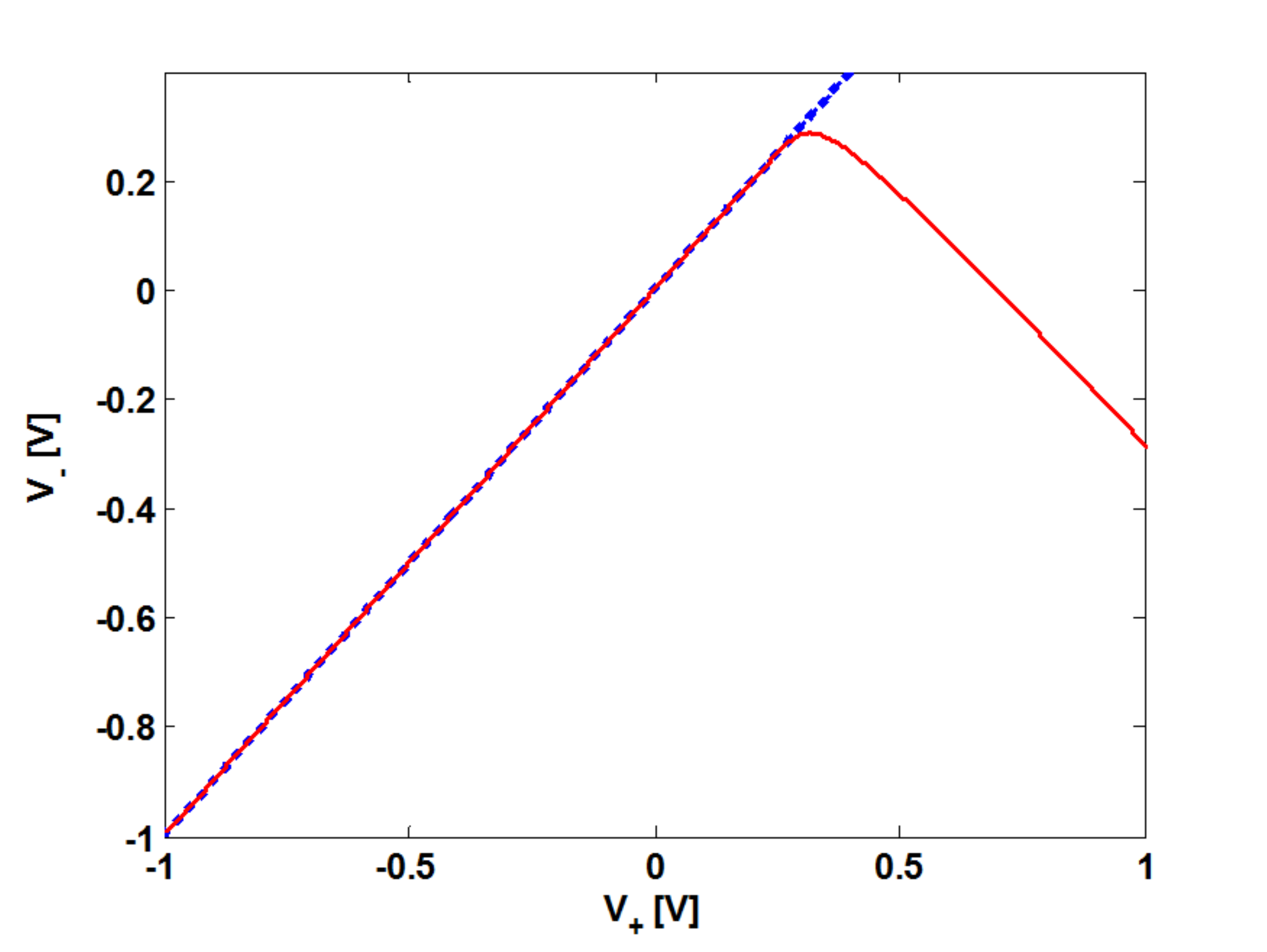}
	\caption{(Color online) A plot of Eq. (\ref{eq:dsol}) (using $I_s = 1 \times 10^{-12} A$, $Y_c = 0.02 S$ and $V_T = 0.030 V$), mapping the incoming voltage $V_+$ to the outgoing voltage $V_-$ after reflection from the model diode. The dotted blue line is $V_- = V_+$, showing the deviation from linearity of the diode (plotted in red solid line).}
	\label{fig:2a}
\end{figure*}

\begin{figure*}     
\includegraphics[width=0.9\textwidth]{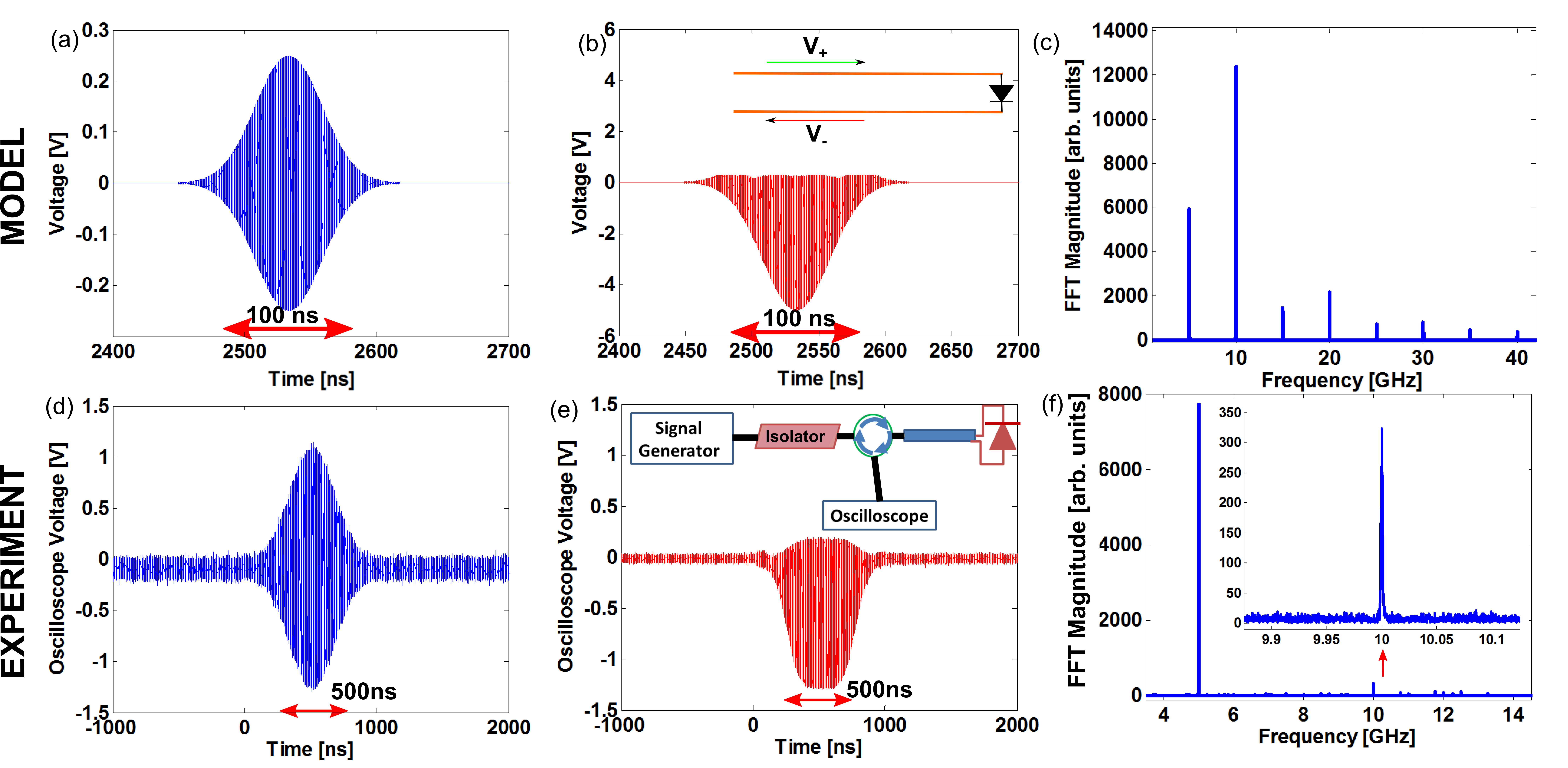}
\caption{(Color online) (a) a low-amplitude Gaussian pulse $V_-$ (100 ns, 5 GHz carrier) reflected from the model diode. (b) a high-amplitude Gaussian pulse $V_-$ (100 ns, 5 GHz carrier) reflected from the model diode. Inset shows a schematic of the model for generating the pulse reflections in (a-c). (c) FFT of the waveform shown in (b). (d) low-power Gaussian pulse (500 ns, 5.0 GHz carrier) reflected from a diode in a single-transmission-line experiment. (e) a high-power Gaussian pulse (500 ns, 5.0 GHz carrier) reflected from a diode in the experiment. Inset shows the experimental setup for generating the pulse reflections in (d-f). (f) FFT of the pulse shown in (e). Inset shows a magnified view of the second harmonic frequency.}\label{fig:2b}
\end{figure*}

Equation (\ref{eq:dsol}) is plotted in Fig. \ref{fig:2a}, using $I_s = 1 \times 10^{-12} A$, $Y_c = 0.02 S$ and $V_T = 0.030 V$. The map between the incoming voltage $V_+$ and the outgoing voltage $V_-$ is similar to an offset tent map.\cite{b1} An important feature of the offset tent map is the small-voltage linearity: for large (positive) incoming voltages, the diode is strongly nonlinear, but for negative, or small positive incoming voltages, the diode is essentially linear. To validate the model diode we consider a simple situation in which a Gaussian pulse of microwave signal is reflected from a diode terminating a single transmission line. In Fig. \ref{fig:2b}(a-b), a Gaussian waveform at a carrier frequency $f_{0} =$ 5 GHz is propagated along a single semi-infinite transmission line terminated in a model diode (as described above), as shown in the inset of (b). For the low-amplitude (0.25V) pulse in (a), the model diode acts as an essentially linear termination, and no nonlinear effect is seen on the reflected pulses. In (b), a high-amplitude (5V) pulse is strongly rectified, creating the reflected pulse shown. The FFT of the rectified pulse is shown in (c), showing several harmonics of the initial signal. 

An experiment was conducted simulating the model transmission line with a coaxial cable terminated by a microwave diode (model 1N4148), and driven by Gaussian pulses from a signal generator, as shown in the inset of Fig. \ref{fig:2b}(e). Reflections of low-power pulses (-35dBm) (show in Fig. \ref{fig:2b}(d) ) are unaffected, while the reflections of high-power (10 dBm) in (e) are strongly rectified. In Fig. \ref{fig:2b}(f), the FFT of the rectified pulse is taken; the inset shows the second harmonic component of the signal. This demonstrates that the diode nonlinearity can be turned ``on'' or ``off'' simply by controlling the amplitude of the signal reaching the nonlinear element. It also demonstrates that the model has the essential property of harmonic generation. The operation of the nonlinear time-reversed mirror depends only on this generic property, and not on any details of the nonlinear mechanism.

\begin{figure*}     
\includegraphics[width=0.5\textwidth]{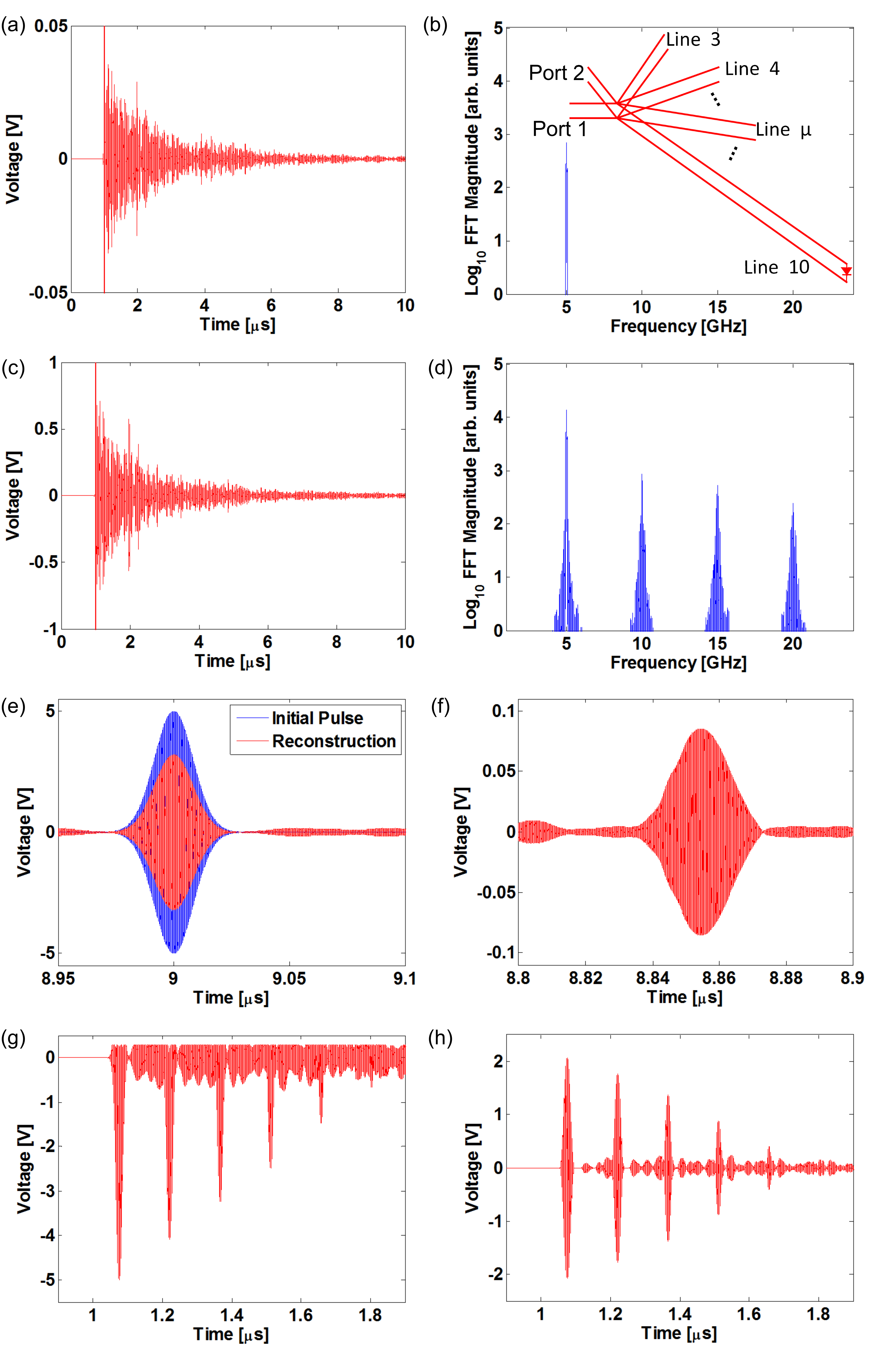}
\caption{(Color online) Waveforms generated by the nonlinear time-reversal star-graph model.  (a) Shows the sona generated by a low amplitude (0.25 V) Gaussian pulse (50 ns, 5 GHz carrier) injected at port 1, recorded at port 2. (b) Shows the frequency spectrum (log FFT magnitude vs. frequency) of the sona in (a), demonstrating the linearity of the network for the low-amplitude pulse. Inset shows a diagram of the star graph network, with the diode terminating the longest line (\#10) (c) Shows the sona generated by a high-amplitude (5 V) Gaussian pulse (50 ns, 5 GHz carrier) injected at port 1, recorded at port 2. (d) Shows the frequency spectrum (log FFT magnitude vs. frequency) of the sona in (c), with harmonics of the initial pulse frequency generated by interaction with the model diode. (e) Reconstruction of the full time-reversed sona in (c) (plotted in red), measured at port 1, compared to the initial Gaussian pulse (blue, larger amplitude). (f) Reconstruction of the time-reversed nonlinear sona (filtered to select second harmonic) measured at the diode location (with the diode removed). (g) Waveform reflected from diode back into the network after the initial excitation. (h) Waveform reflected from diode back into the network after the initial excitation, filtered to select the second harmonic. }\label{fig:modres}
\end{figure*}
\subsection{Full Nonlinear Model}
The nonlinear model of a wave chaotic system is constructed by adding the diode terminating a transmission line to the longest leg of a linear star graph network of transmission lines. The network was constructed from ten line segments of approximate lengths ($1,1,7.07,8.66,11.18,13.23,16.58,18.03,20.61,21.79$)m, 
with no loss ($\alpha_\mu = 0$), and perfectly reflecting ends ($\Gamma_\mu = 1$) for the non-port and non-diode lines. The characteristic admittances were chosen to be $Y_{c,\mu} = 0.02 S$ for lines 3-10, $Y_{c,2} = \sum_{\mu=3}^{n} Y_{c,\mu}$ and $Y_{c,1} = \sum_{\mu=2}^{n} Y_{c,\mu}$.  Port 1 models the linear port of the experiment, while port 2 models the receiving port.  In the time-forward computation, a Gaussian waveform at a fixed carrier frequency $f_{0} =$ 5 GHz is propagated from port 1 into the network, portions propagating along each transmission line (including to the diode). The voltage waves leave the node and propagate down the transmission line to the terminations. For the linear transmission lines, the returning voltage is calculated as a simple reflection (using Eq. (\ref{eq:d})). For the nonlinear line, the outgoing voltage is mapped to the returning voltage via Eq. (\ref{eq:dsol}). Upon returning to the node, the voltage is redistributed amongst the transmission lines. Voltage waves incident on port 1 are absorbed; voltages incident on port 2 are recorded as a sona signal and absorbed. The sona signal contains a linear component (at the original carrier frequency) and a nonlinear component arising from the diode, at several harmonics of the original carrier frequency. 

In the time-reversed computation, the recorded sona signal is band-pass filtered into a linear sona, consisting of only frequencies near $f_0$, and a nonlinear sona, consisting of frequencies near $2f_0$. Separately, the complete sona and each filtered component are time-reversed and propagated into the network (with the diode replaced with a linear reflection) from port 2, and the reconstructions at port 1 and at the diode location are recorded. Figure \ref{fig:modres} shows the various recorded waveforms collected throughout the computation; (a) shows the sona generated at port 2 from a low-amplitude, 50 ns Gaussian pulse (0.25 V) at a carrier frequency of 5 GHz; (b) shows the Fourier transform of the sona, showing the lack of harmonics generated by the diode; (c) is a sona generated from a high-amplitude, 50 ns Gaussian pulse (5 V); while (d) gives the Fourier transform, showing several harmonics of the initial frequency, generated by the nonlinear behavior of the diode. In (e), the entire (unfiltered) high-amplitude sona signal is time-reversed and retransmitted at port 2, reconstructing at port 1. This reconstruction (in red) is overlapped with the time-reversed original Gaussian waveform (blue) showing both a smaller amplitude (from perturbation of the system and loss through the ports) and slight rectification due to the nonlinear components of the sona. (f) shows the reconstruction of the filtered nonlinear sona, measured at the original diode location. This reconstruction occurs at a location in time corresponding to the ballistic propagation distance between the diode and port 1. In (g), the (full frequency spectrum) waveform initially incident upon and reflected from the diode (measured at the diode) is shown, while in (h) the filtered, second-harmonic component of the waveform in (g) is plotted. The strong rectified pulses in (g) (and the corresponding pulses in (h)) represent an initial excitation propagating between the diode and the graph node. This initial excitation  occurs at a time delay corresponding to the ballistic propagation distance between port 1 and the diode, with subsequent pulses delayed by the round-trip length of the line loaded with the diode. The secondary pulses result from an artifact of the model, in which the strong pulse is briefly trapped on the diode-loaded transmission line. Smaller pulses in between these reflected pulses represent signals arriving from other lines in the network, and are weakly rectified compared to the initial ballistic pulse. These results are consistent with the interpretation of the experimental data discussed in section II, that only the first ballistic high-amplitude pulse arriving at the nonlinearity produces most of the harmonic content.

\section{Discussion}
Despite its simplicity, the star graph transmission line model contains the essential features present in the experimental nonlinear time-reversal system, and yields results in good agreement with the experiments. The experimental system contains many more channels for propagation than the model transmission line network, which may account for the quantitative differences in the sona waveforms. Reconstructions of time-reversed sonas in the model are imperfect, as they are in the experiment, due to absorption of signal at the non-receiving port, and due to energy still reverberating in the system past the duration of the sona recording. The model diode simulates a stronger nonlinearity than is used in the single transmission line experiment; the model diode generates several harmonics of the initial signal, compared to the weak second harmonic observed in the experiment. However this detailed difference does not affect the principle of nonlinear time-reversal. The signals created by the nonlinear object propagate through a linear scattering environment before they are recorded. Hence it does not matter what amplitude signal arrives at the nonlinear object as long as it generates harmonic response, or on the details of the nonlinearity, the function of the nonlinear time reversal mirror is generic.

In both the model and the experiment, the addition of more propagation paths limits the power incident on the nonlinearity in the initial part of the time-forward step, and the strength of the harmonics generated. The exclusive reconstruction of the nonlinear signal upon the location of the nonlinearity is demonstrated in the model, consistent with the experiment. The reconstructed nonlinear signal can be seen at the nonlinear element location, at a time delay equal to the propagation time between the nonlinear element location and the initial transmitting port.  This justifies our interpretation that the initial ballistic signal arriving at the nonlinearity is responsible for the bulk of the nonlinear sona in the experiment.

The ability to "find" a nonlinear object and exclusively direct signals to it opens up new applications, such as wireless power transfer with low background power level, precision hyperthermic treatment of tumors with minimal disruption to other tissue, or detecting changes in both the scattering environment and a nonlinear object.

\section{Conclusions}
We have demonstrated time reversal of electromagnetic signals in a system containing a passive, harmonic generating nonlinear element, in which reverse propagation of the time-reversed received nonlinear excitations forms a reconstruction exclusively upon the nonlinear element. Overlapping of multiple sonas allowing resolvable reconstructions of distinct pulses is demonstrated, allowing for exclusive transmission of data by a series of constructed sonas.  The utility of this technique is demonstrated by transmission of images to the nonlinearity by means of overlapping broadcast of single-pulse nonlinear sonas. A model of the nonlinear system is constructed, using a star graph network of transmission lines, with a passive model diode terminating one line as a nonlinear element in the system. This model recovers features seen in the experimental system, and models time-reversal of linear and nonlinear signals in good agreement with the experimental results.

Acknowledgments: This work was funded by the Intelligence Community Postdoctoral Research Fellowship Program (20101042106000), the Office of Naval Research AppEl Center Task A2 (N000140911190), the Office of Naval Research (N000141310474), the Air Force Office of Scientific Research (FA95500710049), and the Maryland Center for Nanophysics and Advanced Materials.


\begin{thebibliography}{5}
\bibitem{pulseoverlap}{Two time-reversed sona signals have been superimposed to create distinct reconstructions of 2 ns pulses as close as one pulse-width apart.}
\bibitem{stargraph}{These lengths, the number of lines and the admittance values are chosen to give sona signals with temporal dynamics similar to those measured experimentally (see Fig. 5).}
\end{thebibliography}
\end{document}